# The techno-politics of crowdsourced Disaster Data in the Smart City

Wolff, E; Muñoz, F.

## Introduction

This article interrogates the techno-politics of crowdsourced data in the study of environmental hazards such as floods, storms, wildfires and cyclones. We highlight some of the main debates around the use of citizen-generated data for assessing, monitoring and responding to disasters. We then argue that, compared to the number of articles discussing the quality of citizen-generated data, little attention has been dedicated to discussing the social and political implications of this kind of practice.

Citizen science is generally understood as the participation of non-scientist citizens in the generation, collection and even analysis of scientific data through frameworks usually established by scientists [1]. Supported by the ubiquitous presence of smartphones and other information and communications technologies (ICTs) that characterise Smart Cities [2], [3], citizen science projects are one of the approaches most used to collectively gather large amounts of data (crowdsourcing). Several authors have been discussing how crowdsourced data is already transforming how cities are managed [4]–[6] but there is little consensus on the political and social implications of mainstreaming this practice for the participating communities.

While this article does not intend to present definitive answers, it outlines inevitable challenges and indicates potential directions for future studies on the techno-politics of disaster data-collection. Within a techno-politics approach, we argue for a model of "political participation" that recognizes citizens providing data to shape cities as equal experts in the production of knowledge and decision-making, rather than external contributors collecting data for formal authorities. This political participation approach, we believe, would decrease the dependence of formal scientific knowledge on citizens' daily-lived experiences, create horizontal collaborations among diverse stakeholders, in terms of respect and recognition, and increase the humanization of marginalized communities, particularly from the Global South.

## The techno-politics of crowdsourced disaster data in the Smart City

In the last few decades, the popularisation of the internet and smartphones drastically modified how information is generated, processed and analysed. A citizen's capacity to capture images, record sounds and sense environmental conditions using smartphones can play an important role in complementing data from multiple other sources through approaches such as crowdsourcing, big data and citizen science [7]–[9].

While a growing body of literature has been investigating this technological transition, the ethical and political implications of the use of citizen-generated data in critical areas of urban governance is still largely understudied [10], [11]. Highlighting this gap in the literature, this opinion piece will interrogate the emerging literature on the techno-politics of citizen-generated disaster data to outline (i) the degree of participation and political engagement allowed by these practices and (ii) the implications of this kind of data-collection to the most disadvantaged social groups.

**Beyond Cheap, Distributed and Frequent Data: The techno-politics of participation**

Over the last five years, the literature of flood management has documented how smartphones have allowed citizens to produce cheap, well-distributed and up-to-date data on disasters [13]–[16]. Often concerned with data validity, publications in the topic refer to the ubiquitous presence of smartphones as a tool that allows central decision-makers to receive data from previously undocumented areas in previously unfeasible timeframes [17].

Most of these publications, however, are underpinned by the implicit assumption that the role played by the contributing citizen is equivalent to that of a sensor. Under this kind of framework, scientists mine data from social media and other applications to support the production of flood models or to inform emergency response [18]. It is important to note that in several of these studies citizens are not even aware that they contributed to a disaster study. While expanding datasets and providing cheap and distributed information on disasters, this kind of participation often ignores the experiential and cultural aspects of disaster data-collection.

Most of the data collected in these studies tends to be analysed and used by authorities and policy-makers coming from formal organizations and academic institutions with no lived knowledge of the everyday realities of vulnerable communities and limited cultural competency. However, as the literature of Cultural Humility argues, people's realities and their experiences cannot be fully understood and solved by scientists, without the experts first recognizing the intersectionality (biases) from where they are analysing the data, such as gender, race, socio-economic status, and culture [19]. Cultural Humility, as a decolonial approach in urban planning, can provide better insights on who should have the power to manage citizen-generated data as well as who should be responsible for the decisions made using crowdsourced data.

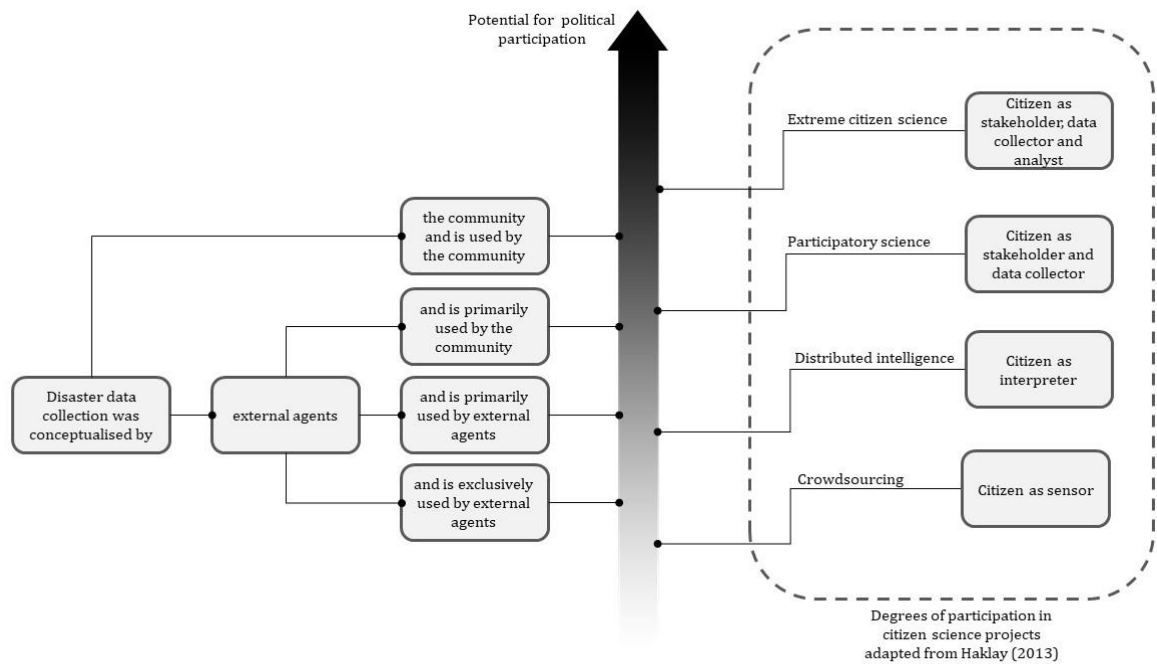

Figure 1: Degrees of political participation in citizen science projects. Left part of the diagram based on Wolff (2021) and right side of the diagram adapted from Haklay (2013).

Increasingly acknowledged as an important element in the social response to disasters, memory and disaster documentation have important political roles that are not activated by most data crowdsourcing projects [20]. Moving beyond the usual concern with data validity, it is useful to review the different types of participation in these projects to explore the political implications of citizen science and crowdsourced data. Several literature reviews on the topic have been discussing the possible roles of citizens in this kind of data-collection framework [21]–[24]. Figure 1 compares Haklay's typology of citizen science projects with the potential of political participation allowed by different projects based on the review of recent publications literature of the field [24].

Even though the model of extreme citizen science might seem theoretical or too "unsystematic" when compared to most data-collection frameworks, we argue that this kind of knowledge production system is already happening in grassroots movements and other community-led initiatives. Experiences in Chile over the last few years exemplify what can be understood as an extreme case of citizen science [25]. In what came to be known as the 2018 social revolution, non-scientist citizens, using grassroots and bottom-up decolonial planning practices, such as 'cabildos públicos' (public gatherings), were able to gather data and use it to organize a socio-economic, political, and environmental agenda without any input from current politicians and scientists. This approach allowed non-scientists to democratically win in a national election the legal opportunity to propose new urban policies and re-write the Chilean constitution of 1980, which was written under the military dictatorship. Chile's case not only informs the current political agenda that authorities are using to manage the country but it has served as a blueprint for extreme social and urban planning changes in other countries in Latin America, such as Colombia.

These examples reveal that emerging frameworks have been allowing citizens to contribute in knowledge production in different roles other than that of just serving as "data-collection devices". Haklay suggests that extreme citizen science and participatory projects have been encouraging citizens to participate as active interpreters and central stakeholders in the disaster management arena [22]. The techno-politics of citizen-generated data, however, are not only determined by the kind of participation allowed by each data-collection framework. In the next subsection, we discuss the promises and the political implications of this kind of knowledge production in historically disadvantaged contexts.

**Where disasters strike the hardest: The techno-politics of crowdsourced disaster data in disadvantaged contexts**

Governments all over the world have faced the challenges of mapping and monitoring hazards in rural and rapidly growing urban areas. This challenge is even more significant in the context of historically disadvantaged groups such as the residents of informal settlements and understudied rural areas in the Global South [26]. In this context, the emergence of social media and the popularisation of smart phones has been seen as a promising game changer. These ICTs have been seen as essential tools to allow governments to better communicate and monitor disasters in urban peripheries and agricultural areas [27], [28]. There is, however, little discussion on how the political and ethical aspects of this framework operate in practice [29].

While the access to smartphones might suggest that disaster information is now more accessible than ever, it is still unclear if ICTs have allowed the most vulnerable residents of cities to participate in discussions that have been historically limited to technical arenas [30], [31]. Some authors argue that smart phones have made internet access ubiquitous and, therefore, completely transformed data-collection and information-sharing before, during and after disasters [31], [32], but others still cast doubts on how accessible these technologies are to the most vulnerable citizens.

Additionally, it is worth considering that informal communities particularly in the Global South, may fear that more information about their location and lives may provide unwanted information to authorities [33], resulting in displacements and the loss of support networks through capital accumulation by land or green dispossession [34], [35]. It is widely accepted that disaster data distribution and frequency are capable of affecting disaster and emergency decision-making [13], [36], [37].

In this realm, other topics also deserve attention such as: how are the most vulnerable going to cope with the additional burden of now serving as data-collectors? And how unequal data availability can affect disaster management? Acknowledging the unequal access to information and communication technologies in historically disadvantaged areas, it is important to consider also how crowdsourcing disaster data could reflect in unequal access to disaster relief services in the Smart City. Similar to how infrastructure networks were framed as political systems by authors in the Science and Technology Studies (STS) [38], studies on disaster politics shows that disaster data-collection practices are competitive tools deeply shaped by political forces that might not account for the voices of the most disadvantaged social groups [30], [39], [40]. In this context, it is imperative to discuss the study of citizen's rights in the Smart City and investigate fair models of data-collection [10], [11].

**Discussion: The need for a broader debate on the Techno-Politics of Disaster data-collection**

The review presented in this article shows that the use of smartphones and other ICTs is quickly gaining traction as a tool to collect data and inform urban planning practices. The literature also suggests that the popularisation of smart phones is expected to facilitate the access to widespread and cheap disaster data in the Smart City, which is particularly important in previously understudied contexts such as in informal settlements and rural areas in the Global South [41], [42]. Beyond facilitating the collection of cheaper and extensive data, this emerging participatory trend is expected to have economic, cultural and political implications that are not yet well understood [10], [11]. As such, it is time for scholars and practitioners to delve into the emerging social and political implications of disaster perception and community-based data-collection in the wake of the Smart Cities.

This article showed that it is imperative to discuss the technical and political implications of citizen-generated data in disaster studies and analyse the intersectionality of those scientists who are analysing the data from their particular points of view. This is the case, for example, when discussing the political implications of this model of disaster management in the context of historically disadvantaged neighbourhoods. Consequently, there are several other questions worth of future explorations such as: Is crowdsourced data technically reliable in terms of characterising disasters? Who owns and who has the power to manage citizen-generated disaster data? Who should be responsible for the decisions made using crowdsourced data?

Evidence suggests that the use of citizen-generated data in disaster studies has the potential to reshape some of the most established technical enclaves of disaster risk management. We argue, therefore, that the discussions on data validity and accuracy that surround citizen science need to make room for a more ample debate on the politics of data collection. While there is little consensus on the definitive role of smartphones in disaster studies, it is clear that the political implications of crowdsourcing disaster data cannot be ignored in any debate about the future of cities.